\documentclass[a4paper,10pt]{article}
\usepackage{rsfs}
\usepackage{latexsym}
\usepackage{amssymb}
\usepackage{graphicx}
\begin{document}

\title{Dispersion Relation for the Nucleon Electromagnetic\\
 Form Factors}
\author{
 Susumu FURUICHI\footnote{Mailing address: Sengencho 3-2-6, Higashikurume Tokyo 203-0012 }\\
 Department of Physics, Rikkyo University, Toshima Tokyo 171-8501, Japan 
 {}\\
 {}\\
 Hirohisa ISHIKAWA\footnote{e-mail address: ishikawa@meikai.ac.jp}\\
 Department of Economy,Meikai University, Urayasu Chiba, 279-8550, Japan
{}\\
{}\\
 Keiji WATANABE\footnote{ e-mail address keijiwatanabe888@yahoo.co.jp; 
 Mailing address: Akazutumi, 5-36-2, 
 Setagaya Tokyo 156-0044}\\
 Department of Physics, Meisei University, Hino Tokyo 191-8506, Japan}
\date{}
\maketitle
\begin{abstract}
  Elastic electromagnetic form factors of nucleons are investigated
 both for the time-like and the space-like momentums by using the 
 unsubtracted dispersion relation with QCD constraints. 
 It is shown that the calculated form factors reproduce the experimental
 data reasonably well; they agree with recent experimental data for the
 neutron magnetic form factors for the space-like data obtained by the 
 CLAS collaboration and are compatible with the ratio of the electric and 
 magnetic form factors for the time-like momentum obtained by the BABAR
 collaboration.x
\end{abstract}

%2009-06-23		
\maketitle

\section{Introduction}

Recently, there are remarkable developments in experiments for the nucleon 
electromagnetic form factors:\\

1)  For the space-like momentum, the ratio of the 
electric and magnetic form factors of proton, $G_E^p$ and $G_M^p$ respectively,  
was shown to be a decreasing function of the 
squared momentum transfer $Q^2$ and the experimental results imply that the 
proton electric form factor vanishes for $Q^2\approx7 ({\rm GeV/c})^2$
 \cite{jonesGE/GM}-\cite{ronGEGM}. \\

2) For the neutron magnetic 
form factor, $G_M^n$, very accurate experimental data 
were obtained and it approximately satisfies $G_M^n(Q^2)/\mu_n \approx
 G_D(Q^2)=(1+Q^2/0.71)^{-2}$, with $Q$ being represented in terms of 
 GeV/c, for fairly wide range of squared momentum transfer 
 $Q^2=1.4-4.8$ (GeV/c)$^2$ \cite{andersionGMn}, 
 \cite{lachnietGMn} (CLAS collaboration).\\

3) For the time-like momentum the ratio $|G_E^p/G_M^p|$ was obtained 
\cite{bardinGptm}, \cite{aubert|GEGM|} (BABAR collaboration), while 
previously the data of form factors had been analyzed under the assumption 
 $G_E^p=0$ or $G_E^p=G_M^p$ .\\
 
Asymptotically, the experimental data of nucleon magnetic factors decrease more 
rapidly than the dipole formula for large $Q^2$ and the decrease has been 
understood as 
a realization of perturbative QCD \cite{brodsky}, the behavior of which can be 
formulated in terms of the dispersion theory with appropriate conditions on the
 absorptive parts; 
 we assumed unsbtracted dispersion relations  for the charge 
 and magnetic moment form factors. To realize the asymptotic form of QCD we
  imposed the superconvergence conditions.
% The imaginary parts are broken up into three  parts; 
% the low, intermediate and asymptotic parts and  the superconvergence 
% conditions are imposed on them.
 
 As the data for the time-like momentum have become accurate, it is necessary 
 to investigate the form factors for the space-like and time-like momentums
  systematically. For this purpose the dispersion theory is effective.
 
The dispersion theoretical calculations performed so far, the value of $G_M^n$ 
turned out to be larger than the above mentioned new experimental data for 
$Q^2=1.4-4.8$ (GeV/c)$^2$ (see Ref. \cite{lachnietGMn}). 
 It is of vital importance to
  investigate if it is 
  possible to realize the experimental data simply by the 
 adjustment of parameters or by the refinement of absorptive parts in the 
 dispersion relation.

 It is the purpose of this paper to analyze experimental data of nucleon form
  factors by 
 the dispersion theory, with the QCD constraints imposed, taking account of the above 
 mentioned new experimental results. 
 
Organization of the paper is given as follows: In Sec.\  2 we explain the 
superconvergent dispersion relation and give conditions
  which are used in this paper. We summarize the absorptive 
parts, which are broken up into three parts: Low, intermediate and asymptotic 
momentum regions. For each momentum region the imaginary parts are given. 
The asymptotic part is expressed as an expansion in terms of the analytically 
regularized running 
coupling constant in the renormalization group for QCD. In Sec.\ 3 we remark
 on
 the numerical analysis. 
In Sec.\ 4 numerical results are summarized. The final section is
 devoted to general discussions.

%%%%%%%%%%%%%%%%%%%%%%%%%%%%%%%%%%%%%%%%%%%%%%%%%%%%%%%%%%%%%%%%%%%%
 \section{Dispersion Relation for the Electromagnetic Form Factors}
%%%%%%%%%%%%%%%%%%%%%%%%%%%%%%%%%%%%%%%%%%%%%%%%%%%%%%%%%%%%%%%%%%%%

 We assume the unsubtracted dispersion relations for the charge and magnetic 
 moment form factors, $F_1^{I}$ and $F_2^I$, respectively, with $I$ 
denoting the isospin state $I=0,1$. That is,
\begin{equation}
 F_i^{I}=\frac{1}{\pi}\int_{t_0}^{\infty}dt'
  \frac{{\rm{Im}}\, F_i^{I}(t')}{t'-t},\quad(i=1,\,\,2) \label{unsubtracted}
\end{equation}
where the threshold is $t_0=4\mu^2$. Here $\mu$ is the pion mass being  
taken as the average of the neutral and charged pion masses. We impose 
conditions on ${\rm{Im}}\, F_i^{I}$ to realize the QCD conditions.

\subsection{Superconvergence Condition and QCD}
Experimental data imply that the magnetic form factors of nucleon decrease 
more rapidly than the dipole formula for large squared momentum transfer.
The decrease agrees with the prediction of perturbative QCD, where magnetic
 form factors of nucleon 
decrease for $Q^2\to \infty$ as
\begin{equation}
 G_M(q^2)\to {\rm const}\frac{\alpha_S(Q^2)^2}{Q^4}
  \left(\ln \frac{Q^2}{\Lambda^2}\right)^{4/3\beta_0},
\end{equation}
where $\alpha_S$ is the running coupling constant of QCD and 
$\beta_0=11-2n_f/3$ with $n_f$ being the number of flavor. $\Lambda $ is the 
QCD scale parameter having the dimension of momentum.

To realize the QCD predictions we impose the following conditions on 
the charge and magnetic moment form factors:
\begin{eqnarray}
 F_1(Q^2)&\to& {\rm const}/[Q^2 (\ln Q^2/{\Lambda}^2)^{\gamma}], \nonumber \\
 F_2(Q^2)&\to& {\rm const}/[Q^4 (\ln Q^2/{\Lambda}^2)^{\gamma}], \label{QCD12}
\end{eqnarray}
for $Q^2 \to \infty$ with $\gamma\ge2$.

We briefly summarize the asymptotic
 theorems which are used to incorporate the 
 constraints of QCD \cite{brodsky}, where the proof is given in 
 Ref.\ \cite{nw1}. 
Let $F(t)$  satisfy the dispersion relation (\ref{unsubtracted}), 
and ${\rm Im}F$ is given as
\begin{equation}
 {\rm Im}F(t') = \frac{c}{[\ln (t'/\Lambda^2)]^{\gamma+1}}
   +O\left(\frac{1}{[\ln(t'/\Lambda^2)]^{\gamma+2}}\right)
\end{equation}
for $t\to \infty$ with $\gamma>1$. Then $F(t)$  becomes 
\begin{eqnarray}
 F(t)&=&\frac{1}{\pi}\int_{t_0}^{\infty}dt'
  \frac{c}{(t'-t)[\ln (t'/\Lambda^2)]^{\gamma+1}}\\ \nonumber
 &\to & \frac{c}{\pi\gamma\ln(|t|/\Lambda^2)]^{\gamma}}  \label{disp1}
\end{eqnarray}
for $t\to \pm \infty$. Generally, when $F(t')$  satisfies
\begin{equation}
t^{\prime\,n+1}{\rm Im}F(t')\,\,\to \frac{c}{[\ln(t'/\Lambda^2)]^{\gamma+1}}
  +O\left(\frac{1}{[\ln(t'/\Lambda^2)]^{\gamma+2}}\right) \label{asymp2}
\end{equation}
for $t' \to \infty$ and the superconvergence conditions
\begin{equation}
 \int_{t_0}^{\infty}dt't^{\prime k}{\rm Im}F(t')=0,\quad k=0,1,\cdots,n,
  \label{asym4}
\end{equation}
 $F(t)$ given by (\ref{unsubtracted}) approaches for 
$t \to \pm\infty$ to the following formula: 
\begin{equation}
 F(t)= \frac{1}{\pi}\int_{t_0}^{\infty}dt'\frac{{\rm Im}F(t')}{t'-t}
     \to \frac{1}{t^{n+1}}\frac{c}{\pi\gamma[\ln(|t|/\Lambda^2)^{\gamma}},
     \label{asym5}
\end{equation}
which can be proved by using (\ref{disp1}) and (\ref{asym4}) 
together with the identity
$$
 \frac{1}{t'-t}=-\frac{1}{t}\Big\{1+\frac{t'}{t}+\cdots
   +\Big(\frac{t'}{t}\Big)^n\Big\}+\frac{1}{t^{n+1}}
   \frac{t^{\prime n+1}}{t'-t}.
$$
Indeed, by using (\ref{asym4}) we have
\begin{equation}
 \int_{t_0}^{\infty}dt'\frac{{\rm Im}F(t')}{t'-t}
  =\frac{1}{t^{n+1}}\int_{t_0}^{\infty}dt'
   \frac{t^{\prime n+1}{\rm Im}F(t')}{t'-t},
\end{equation}
which leads to (\ref{asym5}) as $t^{\prime\,n+1}{\rm Im}F(t')$ satisfies 
(\ref{asymp2}).

To obtain the asymptotic formulas (\ref{QCD12}), therefore, we impose the 
superconvergence conditions  on the imaginary part of form factors, 
${\rm Im}F_i^I(t)$ ($i=1,\,\,2$; $I$ denotes isospin) in the unsubtracted 
dispersion relation (\ref{unsubtracted}):\\
\begin{eqnarray}%}
  \frac{1}{\pi}\int_{t_0}^{\infty}dt'\,{\rm Im}F_1^I(t')
     &=&\frac{1}{\pi}\int_{t_0}^{\infty}dt't'\,{\rm Im}F_1^I(t')=0, 
                                                       \label{sup1}\\ 
  \frac{1}{\pi}\int_{t_0}^{\infty}dt'\,{\rm Im}F_2^I(t') \nonumber
   &=& \frac{1}{\pi}\int_{t_0}^{\infty}dt't'\,{\rm Im}F_2^I(t') 
   = \frac{1}{\pi}\int_{t_0}^{\infty}dt't^{\prime\,2}\,{\rm Im}F_2^I(t')
   =0,                                                       \label{sup2}
\end{eqnarray}%}
where ${\rm Im}F_i^{I}(t')$ satisfies the asymptotic conditions for $t' \to
 \infty$
\begin{equation}
 t^{\prime\, i}{\rm Im}F_i^{I}(t') \to {\rm const}/[\ln (t'/\Lambda^2)]^{\gamma+1}
 \quad (i=1,2).   \label{asymptotic}
\end{equation}

In addition to the conditions (\ref{sup1}) and (\ref{sup2}) we impose the 
normalization conditions at $t=0$:
\begin{eqnarray}%}
 \frac{1}{2} &=& \frac{1}{\pi}\int_{t_0}^{\infty}dt'\,{\rm Im}F_1^{I}(t')/t', 
                                    \label{norm1} \\
   g^I &=& \frac{1}{\pi}\int_{t_0}^{\infty}dt'\,{\rm Im}F_2^{I}(t')/t',
                                    \label{norm2}
\end{eqnarray}%}
where $g^I$ is the anomalous magnetic moments of nucleons with the 
isospin $I$.

%%%%%%%%%%%%%%%%%%%%%%%%%%%%%%%%%%%%%%%%%%%%%%%%%%%%
 \subsection{Imaginary part of the form factors}
%%%%%%%%%%%%%%%%%%%%%%%%%%%%%%%%%%%%%%%%%%%%%%%%%%%%

Let us discuss the imaginary parts of nucleon form factors, which are broken 
up into three parts:
 The low momentum, the intermediate, and the asymptotic regions.

 \subsubsection{Low momentum region}
 
The imaginary parts of the charge and magnetic moment form factors, 
${\rm Im}F_i^{V}$, are given in terms of two pion contribution as follows:
%%%%%%%%%%%%%%%%%%%%%%%%%%%%%%%%%%%%%%%%%%%%%%%%%%%%%%%%%%%%%%%
\begin{eqnarray*}%}
 {\rm Im}[F_1^{V}(t)/e] &=& \frac{m}{2}\frac{(t-4\mu^2)}{4m^2-t} \left(\frac{t-4\mu^2}{t}\right)^{1/2}\\%  \nonumber \\
  &\times& {\rm Re}\Big[M^{*}(t) \Big\{f_{+}^{(-)1}(t) -\frac{t}{4m^2}\frac{m}{\sqrt 2}f_{-}^{(-)1}(t)\Big\}\Big],% \\
\end{eqnarray*}%}
\begin{eqnarray}%}
 {\rm Im}[2mF_2^{V}(t)/e] &=& \frac{m}{2}\frac{(t-4\mu^2)}{(4m^2-t)}
  \left(\frac{t-4\mu^2}{t}\right)^{1/2} \label{ImH}\\%      
   \nonumber\\
   &\times& {\rm Re}\Big[M^{*}\Big\{\frac{m}{\sqrt 2}f_{-}^{(-)1}(t)
      -f_{+}^{(-)1}(t)\Big\}\Big], \nonumber
\end{eqnarray}%}
where $f_{\pm}^{(-)1}(t)$ are helicty
 amplitudes for $ \pi\pi\leftrightarrow N\bar{N}$, $M(t)$ is the pion 
 form factor and $\mu$ is the pion mass.  The superscript $V$ denotes 
 the iso-vector part. 
For the helicity amplitudes we use the numerical values given by H\"ohler and 
Schopper \cite{hoe} and parameterize $M(t)$ according to them.
\begin{equation}
 M(t) = 
  t_{\rho}\{1+(\Gamma_{\rho}/m_{\rho}d)\}
  [t_{\rho}-t-im_{\rho}^2\Gamma_{\rho}(q_t/q_{\rho})^3\sqrt{t}]^{-1},
\end{equation}
where $m_{\rho}$ and $\Gamma_{\rho}$ are the $\rho$ meson mass and width 
respectively and
\begin{eqnarray}
 t_{\rho}&=&m_{\rho}^2,\quad q_{\rho}=\sqrt{t_{\rho}-\mu^2},\quad \\
  d&=& \frac{3\mu^2}{\pi t_{\rho}}\ln\frac{m_{\rho}+2q_{\rho}}{2\mu}
   +\frac{m_{\rho}}{2\pi q_{\rho}}\Big(1-\frac{2\mu^2}{t_{\rho}}\Big).
\end{eqnarray}
The imaginary parts thus obtained are denoted as ${\rm Im}F_i^{H}\,\,(i=1,2)$ 
hereafter. It must be remarked that the $\rho$ meson contribution is included   in the helicity amplitudes of Ref.\ \cite{hoe}. The uncorrelated kaon pair is 
neglected here
 as the effect was estimated to be small \cite{fw2}.

\subsubsection{Intermediate region}

The intermediate states $4\mu^2\le t \le \Lambda_1^2$ are approximated 
by the addition of the Breit-Wigner terms, with the imaginary part 
  parameterized as follow:
\begin{equation}
 {\rm Im}f_R^{BW}(t) = \frac{g}{(t-M_R^2)^2+g^2},
\end{equation}
 where
\begin{equation}
g=\frac{\Gamma M_R^2(M_R^2+t_{res})^3}{t_{res}^2(M_R^2-t_0)^{3/2}}
 \sqrt{\frac{(t-t_0)^3}{t}}\frac{t^2}{(t+t_{res})^3}.
\end{equation}
Here $M_R$ and  $\Gamma$ are the mass and width of resonance, respectively, the threshold $t_0$ is
 $t_0=4\mu^2$ and $t_{res}$ is treated as an 
adjustable parameter. g is introduced to cut-off the Breit-Wigner
 formula.

We write the intermediate part as the summation of resonances
\begin{equation}
 {\rm Im} F_i^{BW,I}=\sum_n a_n^{I,i}f_{nR}^{I}, \label{ImBW}
\end{equation}
where $I$ is the isospin and $n$ is the labeling of resonances (see Table I).
Here the suffix $i$ denotes $i=1,\,\,2$, corresponding to the charge and 
magnetic moment form factors $F_1^N$ and $F_2^N$ ($N$ = n or p). The same formulas for
 $f_{nR}^{I}$ are used for $i=1$ and $i=2$.

%-------------------------------%
 \subsubsection{Asymptotic region}
%------------------------------%
 We express the form factors as power series in the running coupling constant 
  of QCD, $\alpha_S$. To calculate the absorptive part, it is necessary to 
  perform analytic continuation to the time-like momentum. Here we give 
  only the necessary procedure for the analytic continuation of 
  the running coupling constant to the time-like momentum by using the 
  analytic regularization \cite{dok1} \cite{dok2}, as the 
  formulation is given in Ref.\ \cite{nw1}.
  
 Let $\alpha_S(Q^2)$ be the running coupling constant in the renormalization 
 group calculated by the 
perturbative QCD  as the function of the squared momentum $Q^2$ for the
space-like momentum. We use the three loop approximation for 
$\alpha_S(Q^2)$, which is expressed in the Pad\'e form. 
%%%%%%%%%%%%%%%%%%%%%%%%%%%%%%%%%%%%%%%%%%%%%%%%%%%%%%%%%%%%%%%
%\begin{widetext}%	equation}%}	09-07-28add
\begin{equation}
 \alpha_S(Q^2) = \frac{4\pi}{\beta_0}\Big[\ln(Q^2/\Lambda^2)
  +a_1\ln\{\ln(Q^2/\Lambda^2)\} %\nonumber \\
 +a_2\frac{\ln\{\ln(Q^2/\Lambda^2)\}}{\ln(Q^2/\Lambda^2)}
    +\frac{a_3}{\ln(Q^2/\Lambda^2)}+\cdots\Big]^{-1}.  \label{PQCD}
\end{equation}
%\end{widetext}%	equation}%}
$\Lambda$ is the QCD scale parameter, and $a_i$ are expressed in terms of
 the $\beta$ function of QCD,
%%%%%%%%%%%%%%%%%%%%%%%%%%%%%%%%%%%%%%%%%%%%%%%%%%%%%%%%%%%%%%%
\begin{equation}
 a_1=2\beta_1/\beta_0^2, \quad a_2=4\frac{\beta_1^2}{\beta_0^4}, \quad 
 a_3=
  \frac{4\beta_1^2}{\beta_0^4}\left(1-\frac{\beta_0\beta_2}{8\beta_1^2}\right),
\end{equation}
where
%%%%%%%%%%%%%%%%%%%%%%%%%%%%%%%%%%%%%%%%%%%%%%%%%%%%%%%%%%%%%%%
%\begin{widetext}%	equation}%}	09-07-28add
\begin{equation}
 \beta_0 = 11-\frac{2n_f}{3},\quad \beta_1=51-\frac{19n_f}{3},
 \beta_2 = 2357-\frac{5033}{9}n_f+\frac{325}{27}n_f^2
\end{equation}
%\end{widetext}%	equation}%}
with $n_f$ being the number of flavor.
We perform the analytic continuation of the squared momentum to the time-like 
region, $s$, by the replacement in (\ref{PQCD})
\begin{equation}
 Q^2\to e^{-i\pi}s.
\end{equation}

Then $\alpha_S(e^{-i\pi}s)$ becomes complex and is expressed as follows:
\begin{eqnarray}%*}
 \alpha_S(e^{-i\pi}s) &=& \,1/(u-iv)=\frac{u+iv}{D}, \\
           D &=& \,u^2+v^2,
\end{eqnarray}%*}
where $u$ and $v$ are given as
%%%%%%%%%%%%%%%%%%%%%%%%%%%%%%%%%%%%%%%%%%%%%%%%%%%%%%%%%%%%%%%
%\begin{widetext}%	equation}%}	09-07-28add
\begin{eqnarray}%}
 u &=& \ln(s/\Lambda^2)+\frac{a_1}{2}\ln\{\ln^2(s/\Lambda^2)+\pi^2\} 
 \nonumber \\
   &&+\frac{a_2}{\ln^2(s/\Lambda^2)+\pi^2}
         \Big[\frac{1}{2}\ln(s/\Lambda^2)\ln\{\ln^2(s/\Lambda^2)
       +\pi\theta\}\Big]                   \nonumber \\
    &&+\frac{a_3\ln(s/\Lambda^2)}{\ln^2(s/\Lambda^2)+\pi^2}, % \\	2010-01-29 check
\end{eqnarray}%}
%\end{widetext}%	equation}%}	09-07-28add
%%%%%%%%%%%%%%%%%%%%%%%%%%%%%%%%%%%%%%%%%%%%%%%%%%%%%%%%%%%%%%%
%\begin{widetext}%	equation}%}	09-07-28add
\begin{eqnarray}%}
 v &=& \pi+a_1\theta                       \nonumber \\
    &&-\frac{a_2}{\ln^2(s/\Lambda^2)+\pi^2}
        \Big[\frac{\pi}{2}\ln\{\ln^2(s/\Lambda^2)+\pi^2\}
        -\theta\ln(s/\Lambda^2)\Big]       \nonumber \\
    &&-\frac{\pi a_3}{\ln\{\ln^2(s/\Lambda^2)+\pi^2\}},
\end{eqnarray}%}
%\end{widetext}%	equation}%}	09-07-28add
with
\begin{equation}
 \theta=\tan^{-1}\{\pi/\ln(s/\Lambda^2)\}.
\end{equation}
The running coupling constant is given by the dispersion integral 
 both for the space-like and the time-like momentum
\begin{equation}
 \alpha_R(t)=\int_0^{\infty}dt'\frac{\sigma(t')}{t'-t} \label{regular}
\end{equation}
with
\begin{equation}
 \sigma(t')={\rm Im}\alpha_S(e^{-i\pi}s)=4\pi v/\beta_0D.
\end{equation}
$\alpha_R(t)$ represented by (\ref{regular}) is called analytically 
regularized running coupling constant as it has no singular point for 
$t=-Q^2<0$. 
The regularization eliminates the ghost 
pole of $\alpha_S(Q^2)$, given by (\ref{PQCD}), appearing  at
\begin{equation}
 Q^2=Q^{*2}=\Lambda^2 e^{u^{*}},
\end{equation}
where $u^{*}=0.7659596\cdots$ for the number of flavor $n_f=3$.
Calculating (\ref{regular}), we find that $\alpha_R(t)$ is approximately given by the simple formula with the ghost pole subtracted
\begin{equation}
 \alpha_R(Q^2)\approx \alpha_S(Q^2)-A^{*}/(Q^2-Q^{*2})   \label{regular-1},
\end{equation}
where the residue $A^{*}$ is
\begin{equation}
 A^{*}=4\pi\Lambda^2e^{u^{*}}/\Big\{\beta_0
  \Big(1+\frac{a_1}{u^{*}}
  -a_2\frac{\ln u^{*}}{u^{*2}}+\frac{a_2-a_1}{u^{*2}}\Big)\Big\}.
\end{equation}
We use (\ref{regular-1}) as the regularized coupling constant; for the 
time-like momentum we replace $Q^2 \to e^{-i\pi}s$ in (\ref{regular-1}) as 
was mentioned before.

The QCD parts, $F_i^{QCD,\,I}$ ($i=1,2$; $I$ = 0,1) for the squared time-like 
momentum, are written as follows:
\begin{equation}
 F_i^{QCD,\,I}(s)= \hat{F}_i^{QCD,\,I}(s)h_i(s),  \label{qcd1}
\end{equation}
where $\hat{F}_i^{QCD,\,I}$'s are given as expansion in terms of the 
running coupling constant 
\begin{equation}
 \hat{F}_i^{QCD,\,I}(s)=\sum_{j\ge 2}c_j^{QCD,\,I}
  \{\alpha_R(s)\}^j                               \label{qcd2}
\end{equation}
for the time-like squared momentum $s$. We multiply by the function $h(s)$ in 
(\ref{qcd1}) to assure the convergence of the superconvergence 
conditions (\ref{sup1}) and (\ref{sup2}). 
The following formula is assumed for $h_i(s)$:
\begin{equation}
 h_i(s)=\left(\frac{s-t_{Q}}{s+t_1}\right)^{3/2}
       \left(\frac{t_2}{{s+t_2}}\right)^{i+1},
\end{equation}
which may be interpreted as the form factor for $\gamma \to q\bar{q}$ with 
$t_Q$ being the threshold of the quark antiquark pair.
The parameters $t_Q$, $t_1$ and $t_2$ are 
taken as adjustable parameters and will be determined by the analysis of 
experimental data.

For the time-like momentum, we perform the analytic 
continuation of the regularized effective coupling constant $\alpha_R(Q^2)$ 
to $\alpha_R(s)$ through the equation
\begin{equation}
 \alpha_R(s)=\alpha_R(Q^2e^{-i\pi})={\rm Re}[\alpha_R(s)]
    +i\,{\rm Im}[\alpha_R(s)].
\end{equation}
We express the QCD part as the power series expansion in $\alpha_R(s)$
\begin{equation}
 \hat{F}_i^{QCD,I}(s)=\sum_{2\le j}c_{i,j}^{QCD,\,{\rm I}}
  \{\alpha_R(s)\}^j.  \label{cQCD}
\end{equation}
The summation in (\ref{cQCD}) begins in the second order in the effective
coupling constant so as to realize the logarithmic decrease of the nucleon 
form factors.

Imaginary part of (\ref{cQCD}) is obtained to be
%%%%%%%%%%%%%%%%%%%%%%%%%%%%%%%%%%%%%%%%%%%%%%%%%%%%%%%%%%%%%%%
%\begin{widetext}%	equation}%}	09-07-28add
\begin{eqnarray}%}
 &&{{\rm Im}\hat{F}}_i^{QCD,I}
 =\,2c_{i,2}^{QCD,\,{\rm I}}{\rm Re}\,\alpha_R
 {\rm Im}\,\alpha_R  \nonumber \\
  &&\quad+ c_{i,3}^{QCD,\,{\rm I}}[3({\rm Re}\,\alpha_R)^2{\rm Im}\,\alpha_R
   -({\rm Im}\,\alpha_R)^3] \nonumber \\
  &&\quad+c_{i,4}^{QCD,\,I}[4({\rm Re}\,\alpha_R)^3{\rm Im}\,\alpha_R
     -4{\rm Re}\,\alpha_R({\rm Im}\,\alpha_R)^3]   \nonumber \\
  &&\quad+\cdots, \label{ImQCD}
\end{eqnarray}%}
%\end{widetext}%	equation}%}	09-07-28add
and 
\begin{equation}
{\rm Im}F_i^{QCD,\,I}(s)={\rm Im}\hat{F}_i^{QCD\,I}(s)h_i(s).\label{ImQCD1}
\end{equation}

We write the low energy part, intermediate resonance part and asymptotic 
QCD parts of form factors as $F_i^{\rm{ H}}$, $F_i^{BW, I}$ and $F_i^{QCD,\,I}$, 
respectively, which are 
given by the dispersion integral with the imaginary parts (\ref{ImH}),
 (\ref{ImBW}) and (\ref{ImQCD1}). The form factors $F_i^{I}$ 
 are defined by adding them up. We impose the conditions (\ref{sup1}) and 
 (\ref{sup2}) on ${\rm Im}F_i^{I}$ so that the QCD conditions are satisfied.

%%%%%%%%%%%%%%%%%%%%%%%%%%%%%%
 \section{Numerical Analysis}
%%%%%%%%%%%%%%%%%%%%%%%%%%%%%%
We analyzed the experimental data of nucleon electromagnetic form factors 
 $G_M^p/\mu_pG_D$, $G_E^p/G_D$, $G_M^n/\mu_nG_D$ 
$G_E^n$ and the ratio $\mu_p G_E^p/G_M^p$ for the space-like momentum transfer, and 
$|G^p|$ and $|G^n|$ in Refs.\ \cite{bostedGMp}-
\cite{madeyGEn/GMn} for the time-like momentum transfer and the above mentioned 
recent experimental data
  $G_M^n$ for the space-ike and 
$|G_E^p/\mu_p G_M^p|$ for the time-like momentum transfer. 
 The parameters appearing in the 
formulas are determined so as to minimize $\chi^2$.

As was mentioned in the introduction we analyze by taking account of 
the recent experimental data: (a) $G_M^n$ for $Q^2 = 1-4.8$ (GeV/c)$^2$ 
(CLAS collaboration) and (b) $|\mu_pG_E^p|/|G_M^p|$ (BABAR collaboration).
 
In order to see how the situation changes by taking account of these new 
experiments in addition to the other data,
we perform analysis for the following two cases in the $\chi^2$ analysis:\\

Case I: Both of the experimental data, (a) $|\mu_pG_E^p/G_M^p|$ for the 
time-like 
momentum and 
(b) new data for $G_M^n$ for the space-like momentum, are added.\\

Case II: Only the data (a) $|\mu_pG_E^p/G_M^p|$ for the time-like 
momentum are added.\\

Let us remark on the experiments for the time-like momentum 
\cite{bardinGptm}, \cite{ablikimGptm}, \cite{antonelliGntm}, where 
 the form factors 
$|G^p|$ and $|G^n|$ are determined by using the 
formula for the cross section  $\sigma_0$ for the processes 
$e+\bar{e} \to N+\bar{N}$ or $N+\bar{N} \to e+\bar{e}$, which is given as
\begin{equation}
 \sigma_0=\frac{4\pi\alpha^2\nu}{3s}\left(1+\frac{2m_p^2}{s}\right)
  |G(s)|^2.
  \label{exp}
\end{equation}
Here $\alpha$ is the fine structure constant and $\nu$ is the nucleon
 velocity. 
$|G_M^N|$ are estimated from $|G|$ under the assumption $G_M=G_E$ or $G_E=0$. 
$\sigma_0$ is expressed in terms of $G_M^N$ and $G_E^N$ as follows:
\begin{equation}
 \sigma_0=\frac{4\pi\alpha^2\nu}{3s}
 \left(|G_M^N|^2+\frac{2m^2}{s}|G_E^N|^2\right). \label{theory}
\end{equation}
Equating (\ref{exp}) and (\ref{theory}), we have
\begin{equation}
 |G|^2=\frac{|G_M^N|^2+2m^2|G_E^N|^2/s}{1+2m^2/s}.  \label{timedata}
\end{equation}
Substituting our calculated result of form factors to the right hand side 
of (\ref{timedata}), we obtain 
 the theoretical value for $|G|$, which is compared with the experimental 
data for the magnetic form factor obtained under the assumption $G_M=G_E$.

The parameters appearing in our analysis are the following: 
Residues at resonances, coefficients appearing in the expansion by the 
QCD effective coupling constants, cut-offs for the intermediate region 
$\Lambda_1$.
In addition to them  we have parameters in the Breit-Wigner formula and 
the convergence factor $h$ of QCD contribution, $t_Q$, $t_{res}$, $t_1,\,\,
t_2,\,\, t_3$.

We have taken the masses and the widths of resonances as adjustable parameters.
As the superconvergence constraints impose very stringent conditions on the 
form factors, it was necessary to take the masses and widths as parameters. \\

%%%%%%%%%%%%%%%%%%%%%%%%%%%%%%
 \section{Numerical Results}
%%%%%%%%%%%%%%%%%%%%%%%%%%%%%%

 We give in Tables 1, 2 and 3 the results for the parameters 
 for the cases I and II obtained by the $\chi^2$ analysis; in Table 1 the masses 
and widths of 
 resonances and in Table 2 residues at 
 resonance poles and in Table 3 the coefficients
 $c_{i,j}^{QCD,\,I}$ $(i=1,2; \,\,j=2,3,4;\,\,I=0,1)$
 in the expansion in terms of the
 effective coupling constant $\alpha_R$ of QCD defined by (\ref{cQCD}). 
 The number of flavor is taken as  $n_f = 3$. ${\rm Im} F_i^H$ is cut-off at 
 $\Lambda_0^2=0.779$
 GeV$^2$ and the Breit-Wigner formulas at  $\Lambda_1$ = 26.0 GeV.
 The QCD parameter is fixed at $\Lambda$ = 0.216 GeV. The other parameters 
 are determined as follows:\\
 Case I: $t_0$ = 4$\mu^2$, $t_1$ = 0.243$\times 10^3$ GeV$^2$, 
 $t_2$ = 0.237$\times 10^3$ GeV$^2$, $t_{res}$ = 0.2260$\times10^3$ GeV$^2$,
  $t_Q$ = 0.202$\times 10^2$ GeV$^2$.\\
 Case II: The same as in the case I except for $t_{res}$ = 0.2253$\times10^3$ 
 GeV$^2$. 
 
  The value of 
 $\chi^2$ is obtained to be $\chi^2_{tot}=393.4$ for the case I and 
 $\chi^2_{tot}= 308.7$ for the case II, which includes both 
 the data of space-like and time-like regions. 
 The total number of data is 245 for the Case I and 236 for the 
 Case II. Number of parameters is 36 so that DOF/$\chi_{min}$ = 1.88 for the 
 Case I and 1.54 for the Case II.
 
%%%%%%%%%%%%%%%%%%%%%%%%%%%%%%%%%%%%%%%%%%%%%%%%%%%%%%%%%%%%%%%
\begin{table}
 \caption{Masses and widths determined by the $\chi^2$ analysis for
 the cases I and II.\hspace{64pt}$\quad$}
\begin{tabular}{cc|cl|ccll}\hline
 {}& {}& $\hspace{20pt}$ case I &{} & $\hspace{20pt}$ case II&{}\\  \hline
{}&{}&  mass & width  & mass & width \\ 
isospin & $n$  &(GeV/c$^2$)&(GeV)&(GeV/c$^2$)&(GeV) \\\hline
   {} & 1 & 1.341 & 0.3221 & 1.352 & 0.325 \\
   {} & 2 & 1.379 & 0.2204 & 1.370 & 0.220 \\
$I=1$ & 3 & 1.599 & 0.2636  & 1.587 & 0.264 \\
   {} & 4 & 1.824 & 0.3679 & 1.826 & 0.368 \\
   {} & 5 & 2.048 & 0.3848 & 2.100 & 0.398 \\ \hline
   {} & 1 & 0.78256 &  0.844$\times 10^{-2}$ & 0.78256
            &  0.844$\times 10^{-2}$ \\
   {} & 2 & 1.01945 &  0.426$\times 10^{-2}$ & 1.01945 &  0.426$\times 10^{-2}$ \\
$I=0$ & 3 & 1.212 & 0.1582 & 1.206 & 0.1584 \\
   {} & 4 & 1.437 & 0.2102 & 1.440 & 0.2104 \\
   {} & 5 & 1.505 & 0.1281 & 1.510 & 0.1285 \\ \hline
\end{tabular}
\end{table}
\begin{table}
\caption{The coefficients $a_i^{I,n}$, residues at the resonance poles, 
determined by the $\chi^2$ analysis for the cases I and II.}
\begin{tabular}{cc|cc|cc}\hline
 {} & {} & $\hspace{20pt}$ case I & {} & $\hspace{20pt}$ case II {} \\ \hline
 isospin & $n$ & $a_1^{I,n}$(GeV$^2$)&$a_2^{I,n}$(GeV$^2$) &
                 $a_1^{I,n}$(GeV$^2$)&$a_2^{I,n}$(GeV$^2$) \\ \hline
 {} & 1 & $-$4.66     &  8.45    & $-$4.47 & 8.37  \\
 {} & 2 & $\,\,\,\,$8.489277 &$\,-$17.50252 &$\quad\,$7.4739945 &$-$15.8900\\ 
$I=1$ & 3 & $\,-$9.623356 &$\,\,\,\,$ 13.46278 & $-$8.050218 &
$\quad\,$10.93951  \\
 {} & 4 & $\,\,\,\,$ 7.065036 &$\,\,-$7.310033 &$\quad$6.091668 & 
 $\quad\,-$6.071918 \\
 {} & 5 & $-$0.140    & 1.36     & $-$0.118    &   1.11     \\ \hline
 {} & 1 &$\quad$0.899887 & 0.02568286 &  0.8762127 & 0.09468219 \\
 {} & 2 & $-$3.625433  & 0.5913331 & $-$3.514823  &$\,\,\,\,$0.3151743 \\
$I=0$ & 3 &$\,\,\,\,$7.385961  &$-$2.033127 &$\,\,\,\,\,$6.954618  & 
$-$1.526721 
  \\
 {} & 4 & $-$3.934473  & $-$1.019970 &  $-$3.579443  & $-$2.125879 \\
 {} & 5 & $-$1.028184  & 2.582630 &  $-$1.038278  &$\,\,\,\,\,$3.394527  
 \\ \hline
\end{tabular}
\end{table}

\begin{table}
\caption{The coefficients $c_{i,j}^{QCD,I}$ of the QCD terms for the cases 
I and II determined by the $\chi^2$ analysis.}
\begin{tabular}{cc|cll}\hline
  {} & {} & {} &case I & {}  \\ \hline
isospin & $i$ & $c_{i,2}^{QCD,I}$ & $c_{i,3}^{QCD,I}$ & $c_{i,4}^{QCD,I}$ \\
    \hline
$I=1$ & 1 &$\,\,$ 0.5505731 & $-$4.12 & $-$6.50 \\
  {} & 2 & 3.758361  & $-$0.4002$\times 10^2$ & 0.6224$\times 10^2$ \\ \hline
$I=0$ & 1 & $\,\,$1.108707  & $-$2.76  & $-$0.7045 $\times 10^2$ \\
  {} & 2 & $-$5.215940 & 0.6706$\times 10^2$ & $-$0.19908$ \times 10^3$ \\ 
  \hline
   \hline
   {} & {} & {} & case II & {} \\ \hline
 isospin & $i$ & $c_{i,2}^{QCD,I}$ & $c_{i,3}^{QCD,I}$ & $c_{i,4}^{QCD,I}$ \\
    \hline
$I=1$ & 1 & $-$0.7186148$\times 10^{-1}$ & $\quad$1.48 & $-$6.99  \\
  {} & 2 &  4.252983   & $-$0.4375 $\times 10^2$ & 0.5543 $\times 10^2$ \\ 
     \hline
$I=0$ & 1 &$\quad$ 0.8918455  & $-$1.10  & $-$0.6787000 $\times 10^2$ \\
 {} & 2 & $-$5.617625  & 0.7029$\times 10^2$  & $-$0.19551$ \times 10^3$ \\
   \hline
\end{tabular}
\end{table}

We illustrate in Figs.\ 1 - 9 the calculated results for the form factors. 
The results 
for the Case I is given by the solid curve and Case II by the dashed one. 
Figs.\ 1 - 4 the results for the space-like momentum are illustrated: Fig.\ 1 
the proton magnetic form factors $G_M^p/\mu_pG_D$, Fig.\ 2 proton electric 
form factor $G_E^p/G_D$, Fig.\ 3 the neutron magnetic form factor 
$G_M^n/\mu_n$ and Fig.\ 4 the neutron electric form factor. In Fig.\ 5 we 
illustrate the ratio of proton electric and proton magnetic form factors
 $\mu_pG_E^p/G_M^p$. We find that $G_E^p=0$ at $Q^2=6.57$ (GeV/c)$^2$ for the 
 case I and $Q^2=6.79$ (GeV/c)$^2$ for the case II. The form factor for the 
time-like momentum $|G|$  is
 given in Fig.\ 6 for the proton and in Fig.\ 7 for the neutron. The result for
 the proton form factor agrees with the experimental data, but for the neutron 
 the calculated one becomes larger than the experiments for large $Q^2$.

In Fig.\ 8 we compare the calculated result for the neutron magnetic form 
factor $G_M^n/\mu_nG_D$ with the recent experiments. The solid curve agrees 
with the experimental data very well. The dashed one becomes a little 
larger than the result obtained by the CLAS collaboration. However, the 
deviation is not very large. In Fig.\ 9 we illustrate the result for 
$|G_E^p/\mu_pG_M^p|$ for the time-like momentum. There seems to be some 
discrepancy between the experimental data: The ratio obtained by 
Bardin el al. \cite{bardinGptm} is smaller than that of Aubert et al.
 \cite{aubert|GEGM|}. Our result coincides with the result of Bardin et al. 
for small $Q^2$ and that of Aubert el al. for large $Q^2$. 

\begin{figure}
\begin{center}
\includegraphics[width=.5\linewidth]{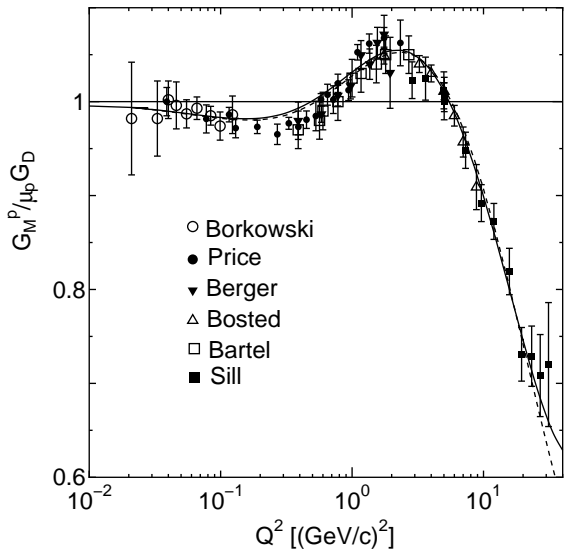}
\end{center}
 \caption{Proton magnetic form factor for the space-like momentum. The solid 
 curve is the result for case I and the dashed one for the case II.}
\end{figure}%

\begin{figure}
\begin{center}
\includegraphics[width=.5\linewidth]{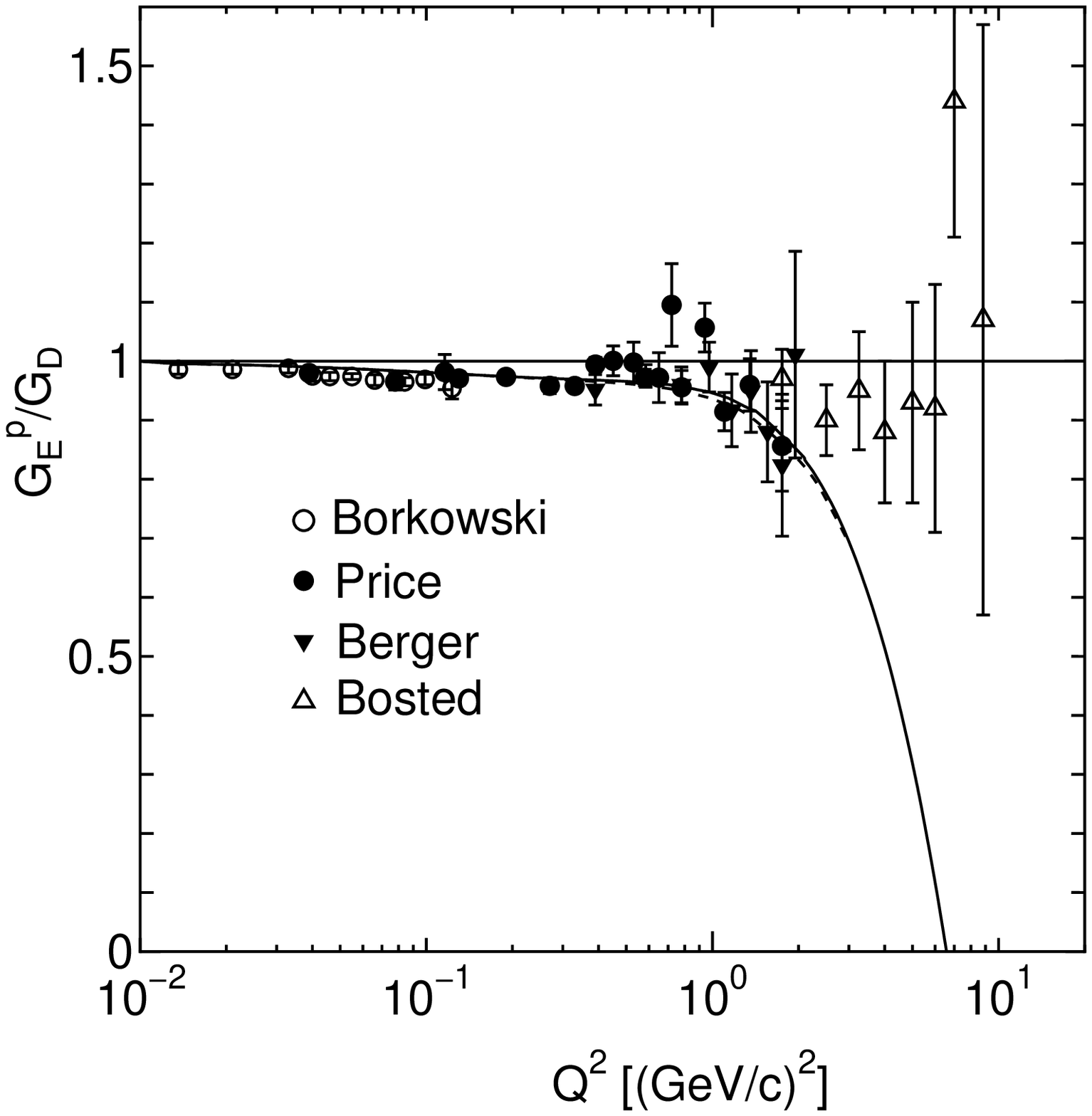}
\end{center}
\caption{Proton electric form factor for the space-like momentum. The solid 
 curve is the result for case I and the dashed one for the case II.}
\end{figure}%

\begin{figure}
\begin{center}
 \includegraphics[width=.5\linewidth]{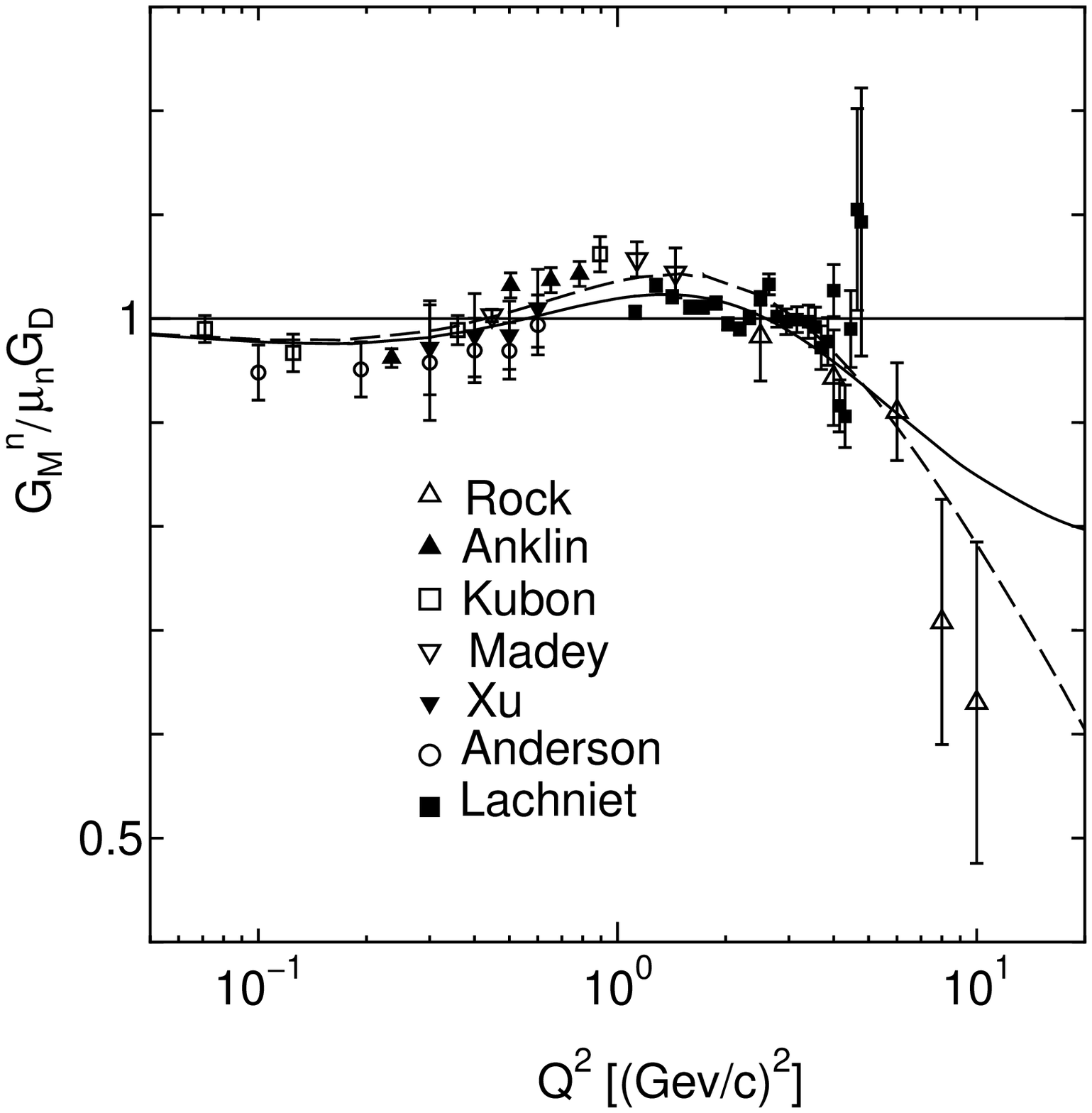}
\end{center}
\caption{Neutron magnetic form factors for the space-like momentum. The solid 
 curve is the result for case I and the dashed one for the case II.}
\end{figure}%

\begin{figure}
 \begin{center}
  \includegraphics[width=.5\linewidth]{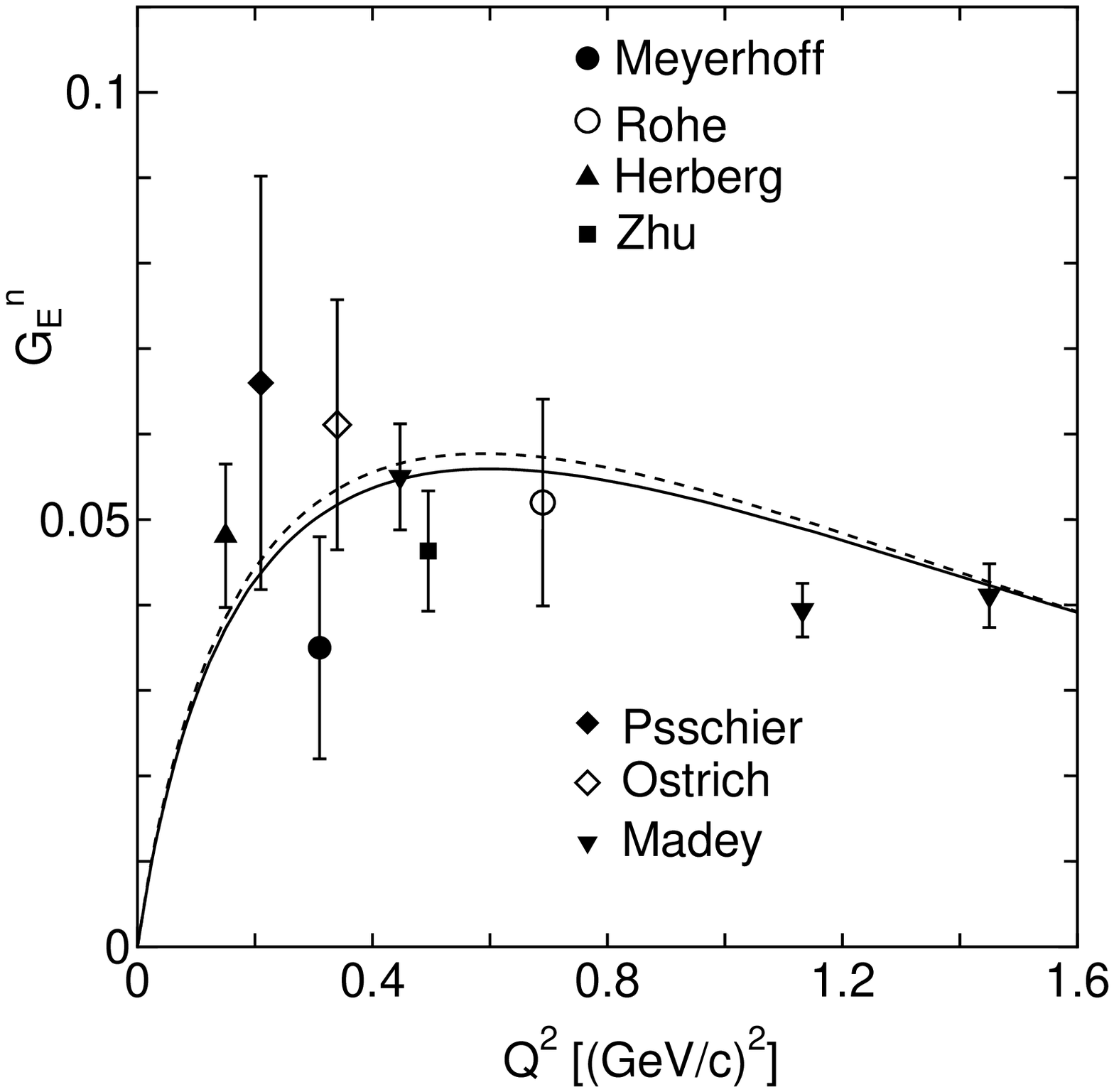}
 \end{center}
\caption{Neutron electric form factor for the space-like momentum. The solid 
 curve is the result for case I and the dashed one for the case II.}
\end{figure}%

\begin{figure}
\begin{center}
\includegraphics[width=.5\linewidth]{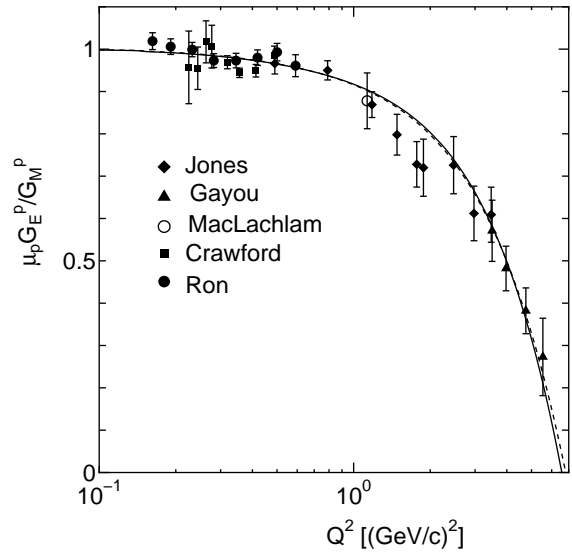}
\end{center}
\caption{ Ratio of the electric and magnetic form factors of proton for the 
 space-like momentum. The solid 
 curve is the result for case I and the dashed one for the case II.}%
\end{figure}

\begin{figure}
 \begin{center}
  \includegraphics[width=.5\linewidth]{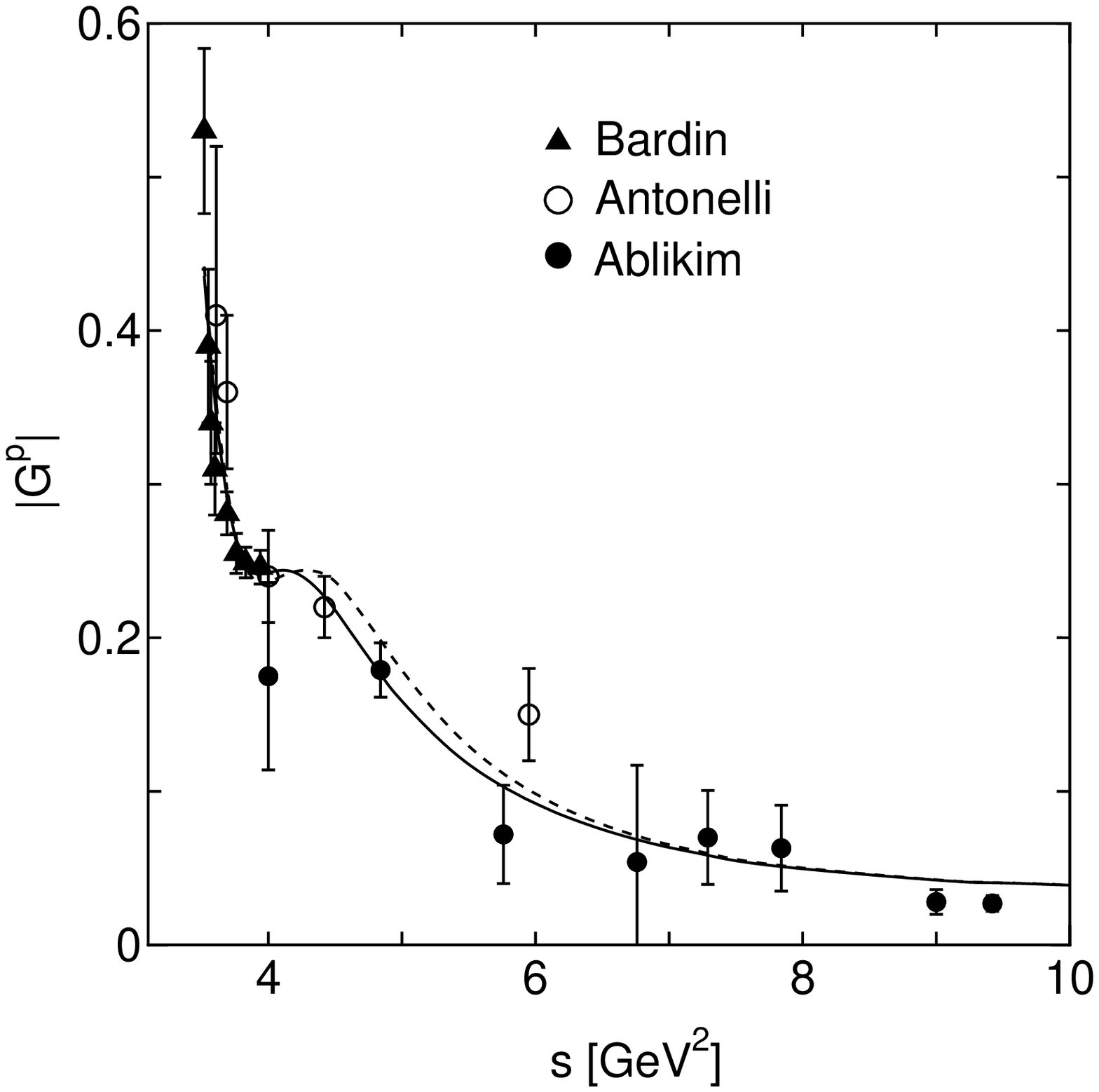}
 \end{center}
  \caption{Proton form factors for the time-like momentum. The solid 
 curve is the result for case I and the dashed one for the case II.}
 \end{figure}%
\begin{figure}
 \begin{center}
  \includegraphics[width=.5\linewidth]{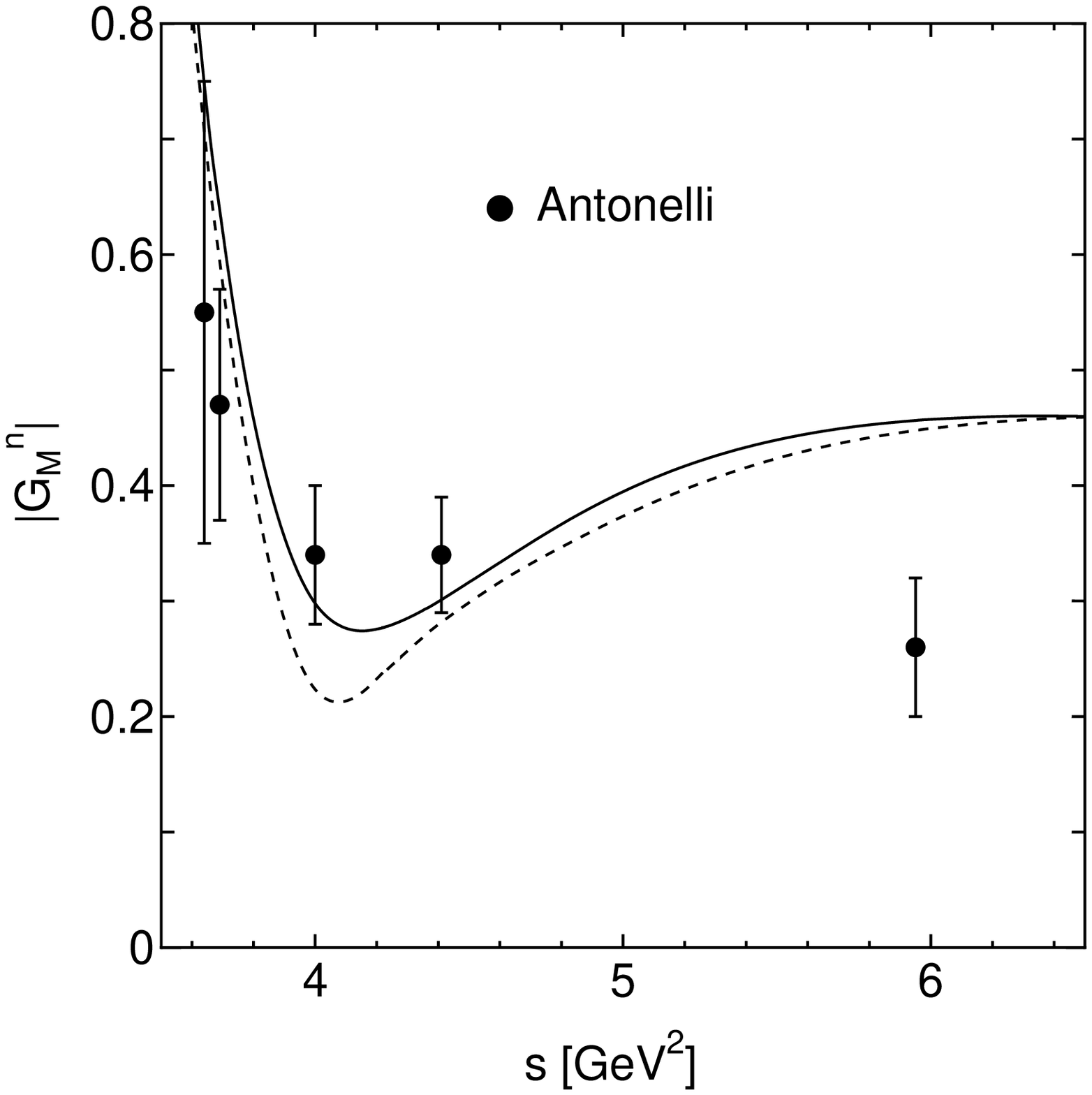}
 \end{center}
\caption{The neutron form factor for the time-like momentum. The solid 
 curve is the result for case I and the dashed one for the case II.}
\end{figure}%
\begin{figure}
 \begin{center}
  \includegraphics[width=.5\linewidth]{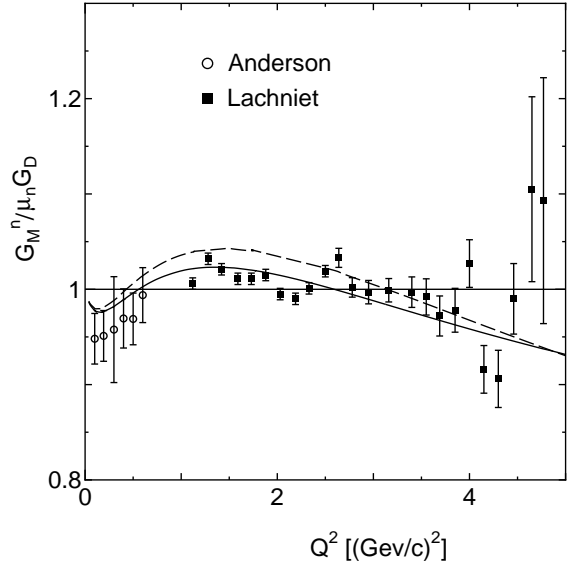}
 \end{center}
  \caption{Neutron magnetic form factor for the space-like momentum 
   in the few (GeV/c)$^2$ region. The solid 
 curve is the result for case I and the dashed one for the case II.}
 \end{figure}%
\begin{figure}
 \begin{center}
  \includegraphics[width=.5\linewidth]{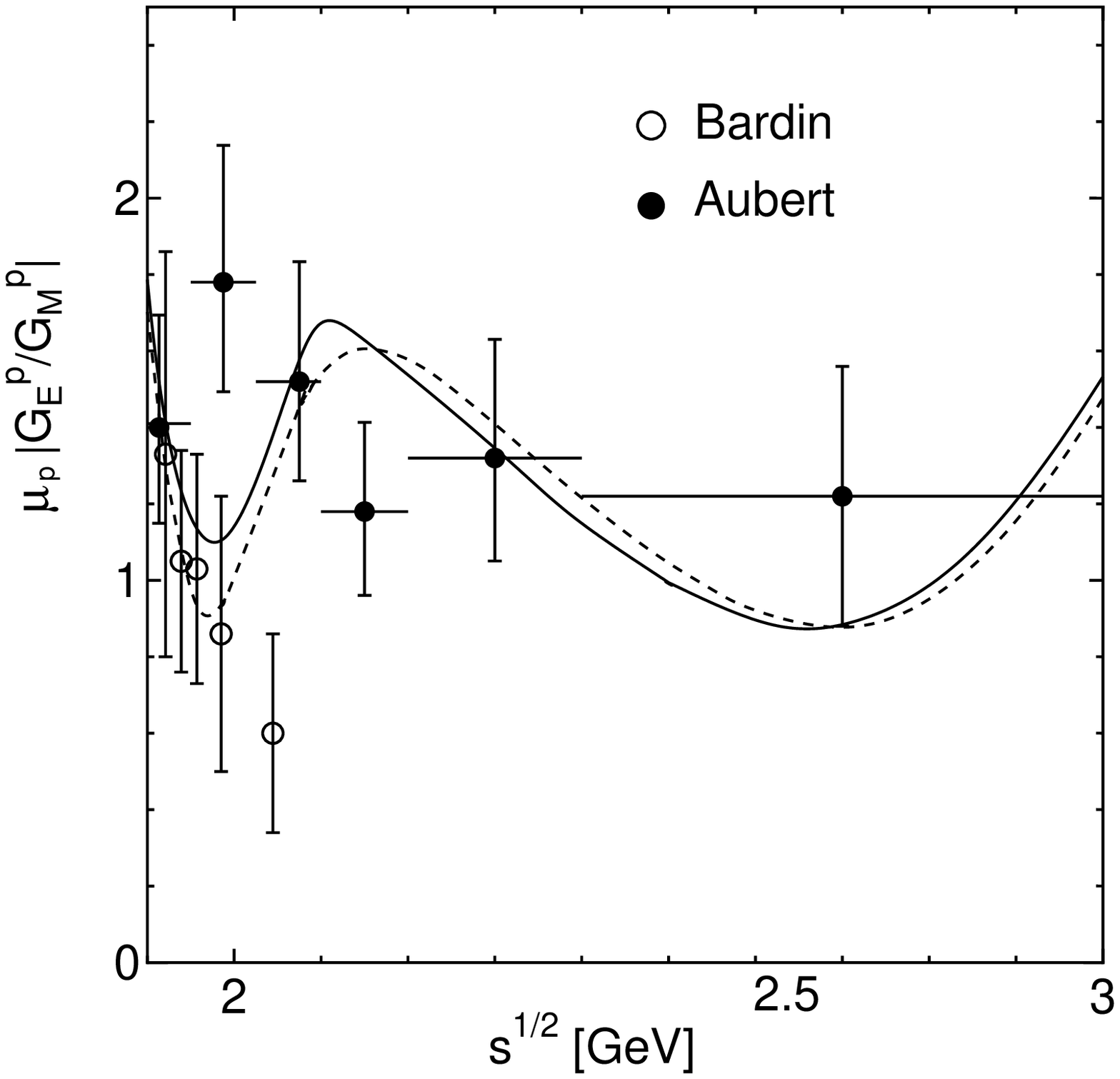}
 \end{center}
\caption{$|G_E^p/\mu_pG_M^p|$ for the time-like momentum. The solid 
 curve is the result for case I and the dashed one for the case II.}
\end{figure}%

%%%%%%%%%%%%%%%%%%%%%%%%%%%%
\section{Concluding Remarks}
%%%%%%%%%%%%%%%%%%%%%%%%%%%%

  The experimental data for the neutron magnetic form factor for the
  space-like momentum with $Q^2=1.4 - 4.8$ (GeV/c)$^2$ \cite{lachnietGMn}, 
  mentioned in Sec.\ 1, are reproduced very well by our calculation. 
  
 The absorptive parts of the form factors for the asymptotic region 
 are approximated by the 
 power series in the effective coupling constant of QCD, which begins 
 $O(\alpha_R^2)$ as is given in  (\ref{qcd2}). We have taken three terms in 
 the expansions; the terms of order up to $O(\alpha_R^4)$ are 
 necessary to reproduce the experiments as in the case of deep inelastic 
 electron scattering processes. 
 
It is remarked here that the electromagnetic form factor of bosons,  both 
for the space-like and time-like momentums, can be explained with recourse 
to the superconvergent dispersion relation with the QCD constraints  \cite{nw1}. 
 
For the electric form factor of proton there are deviation of the dispersion 
theoretical calculation from the 
experimental data for large $Q^2$, where  the data  were obtained 
by using the Rosenbluth formula. The discrepancy may imply the necessity of
 correction 
of two photon processes to the experimental data 
 \cite{borisyuk} \cite{belushkin}. 

We used the experimental data for the helicity amplitudes obtained by 
H\"oher and Schopper in which the contribution from the $\rho$ meson 
is included. As their data are limited to low $t \,\,(\le 0.779$ (GeV/c)$^2$), 
we do not have sufficient 
data for the region $s \le 4m_N^2$. We 
supplemented the unphysical region for $I=1$ state by introducing vector bosons
 with the mass, $m_V \stackrel{\large <}{_{\sim}}1.4$ GeV/c$^2$.   For the isoscalar state 
we also introduced a vector boson with the  mass about $1.2$ GeV/c$^2$.

In our calculation we treated all of the vector boson masses and widths as 
parameters. If they are kept 
at experimental values, we get poor results. The superconvergence 
conditions are so strong  that the value of $\chi^2$ is very sensitive
 to the mass and width. The masses are obtained to be smaller 
than the experimental value and the existence of vector bosons with the masses 
around 1.2 $\sim$ 1.4 GeV/c$^2$ are necessary both for the $I = 1$ and 
$I = 0$ states.

To conclude the paper we remark on the  mass around 1.2 GeV/c$^2$. We have 
introduced the vector boson to supplement the lack of information on the 
the small $Q^2$. However, both 
  for $I=0$ and $I=1$ states there are indications of resonances observed 
  by the processes $e^{+}e^{-}\to \eta \pi^{+}\pi^{-}$, 
   $\gamma p \to \omega \pi^{0} p$ and $B \to D^{*}\omega\pi^{-}$ \cite{komada}. 
Incorporation of further resonances may improve results for 
   the time-like momentum. \\
   
The authors wish to express gratitude to Professor M. Ishida for the valuable
 discussions  and comments. We also would like to thank Dr. T. Komada for the 
 information on the vector bosons  with the mass around 1.2 GeV/c$^2$.

\end{document}